\documentclass[twocolumn,letterpaper,table]{aastex6}

\usepackage{xcolor}
\usepackage{natbib}
\usepackage{amsmath}
\usepackage{savesym}
\savesymbol{tablenum}
\usepackage{siunitx}
\restoresymbol{SIX}{tablenum}
\usepackage{txfonts}
\usepackage{mathrsfs}
\usepackage{epstopdf}
\usepackage{graphicx,color}
\usepackage{enumerate}
\usepackage{float}
\usepackage{hyperref}

\usepackage{etoolbox}

\newcommand{\rH}{r_{\rm H}}

\newcommand{\Mp}{M_{\rm p}}
\newcommand{\Mj}{M_{\rm J}}
\newcommand{\Mstar}{M_{\ast}}
\newcommand{\Me}{M_{\oplus}}

\newcommand{\secref}[1]{\S\ref{#1}}
\newcommand{\figref}[1]{Figure \ref{#1}}

\begin{document}

\title{PDS~70: A transition disk sculpted by a single planet}

\author{Dhruv Muley\altaffilmark{1}, Jeffrey Fung\altaffilmark{1,3}, Nienke van der Marel\altaffilmark{2}}
\altaffiltext{1}{Department of Astronomy, University of California, Campbell Hall, Berkeley, CA 94720-3411}
\altaffiltext{2}{Herzberg Astronomy \& Astrophysics Programs, National Research Council of Canada, 5071 West Saanich Road, Victoria BC V9E 2E7, Canada}
\altaffiltext{3}{NASA Sagan Fellow}

\email{email: dmuley@berkeley.edu}

\begin{abstract}
The wide, deep cavities of transition disks are often believed to have been hollowed out by nascent planetary systems. PDS~70, a ${\sim}5$ Myr old transition disk system in which a multi-Jupiter-mass planet candidate at 22 au coexists with a ${\sim}30$ au gas and ${\sim}60$ au dust-continuum gap, provides a valuable case study for this hypothesis. Using the \texttt{PEnGUIn} hydrodynamics code, we simulate the orbital evolution and accretion of PDS~70b in its natal disk. When the accreting planet reaches about 2.5 Jupiter masses, it spontaneously grows in eccentricity and consumes material from a wide swathe of the PDS~70 disk; radiative transfer post-processing with \texttt{DALI} shows that this accurately reproduces the observed gap profile. Our results demonstrate that super-Jupiter planets can single-handedly carve out transition disk cavities, and indicate that the high eccentricities measured for such giants may be a natural consequence of disk-planet interaction.

\end{abstract}

\keywords{accretion, accretion disks --- methods: numerical --- planets and satellites: formation --- protoplanetary disks --- planet-disk interactions}

\section{Introduction}
\label{sec:intro}
A super-Jupiter companion, PDS~70b, has recently been directly imaged by SPHERE around the pre-main sequence star PDS~70 \citep{Muller18}. It is located about 22 au away from its host star \citep{Keppler18,Wagner18}, and is inside a transition disk cavity that stretches to {$\sim$}60 au in dust continuum \citep{Hashimoto12,Dong12,Hashimoto15,Muller18,Keppler19}. Could PDS~70b be the cause of this cavity? If so, what can we tell about its properties and evolutionary history from its interaction with the disk?

Transition disk cavities have long been suspected to be emptied by gravitational disk-planet interaction. Planets repel material from their orbits by exerting Lindblad torques, and so open gaps in their disks. Early work by \citet{Quillen04} and \citet{Varniere06}, for example, investigate how planetary gaps may explain transition disks; and \citet{Kley00} showed that with multiple planets, their gaps can merge into a wider, single gap. These early simulations focused on the clearing of the inner disk, which \citet{Crida07} demonstrate to be dependent on the choice of the inner boundary condition. In this study, we turn our attention to the outer disk.

One of the biggest puzzles that the PDS~70 system presents is that PDS~70b's location at 22 au appears too close in to explain a 60 au cavity. Past studies have implied that single-planet gaps are narrow, only a fraction of the orbital radius of the planet \citep[e.g.,][]{Crida06,Fung14,Duffell15,Kanagawa16,Ginzburg18}. Part of the discrepancy lies in how gap sizes are observed. Sub-mm continuum emission  \citep[e.g.,][]{Hashimoto15,Long18} probes only large dust grains, which concentrate at the gap's outer pressure maximum and leave interior regions strongly depleted. However, the distribution of gas itself \citep{vanderMarel16,Dong17}, as well as that of smaller grains (which are more strongly coupled to the gas), can show a much gentler decline. This is indeed the case in PDS~70, where the small-grain cavity, traced by near-infrared scattered light, reaches $\sim$54 au \citep[][]{Hashimoto15,Keppler18,Keppler19}, while the gas gap extends only to $\sim$33 au \citep{Long18}. But even accounting for this, simulations by \citet{Keppler19} find that even a 10 Jupiter-mass ($\Mj$) PDS~70b, on a fixed, circular 22 au orbit, would be incapable of reproducing the observed gap in the system.

One potential explanation for the gap's width is that there exists a second planet lying between 22 and 60 au---a scenario that would resemble, for instance, the simulations of \citet{Zhu2011}, \citet{DodsonSalyk11}, or \citet{DuffellDong15}. On the other hand, it may not be necessary to invoke an unseen companion if one accounts for additional physical processes---in this study in particular, we probe the effects of planetary gas accretion and eccentricity excitation. Both processes are by no means new ideas. One would intuitively expect the former to deepen gaps and the latter to widen them, but whether they are sufficient to quantitatively explain transition disks, or PDS 70 in this case, has not before been studied in detail. The interplay between them and its effects on gap morphology is even less understood.

There are reasons to believe that these processes have played a role in shaping the evolution of the cavity in PDS 70. In terms of accretion, H$\alpha$ observations of PDS 70b with \mbox{MagAO} \citep{Wagner18} provide evidence of ongoing planetary growth. Evolutionary models place the planet's mass between 2 and 17 $\Mj$ \citep{Muller18}, while disk modeling predicts the disk's gas mass to be about 7 $M_{\rm J}$ \citep{Long18}. Therefore, we expect that accretion by PDS~70b has deepened and widened the gap in the system.

The estimated mass of PDS~70b makes it a likely subject of eccentricity excitation by disk-planet interaction. ``Co-orbital'' eccentric Lindblad resonances, those concentrated near planetary orbits\footnote{These resonances would lie exactly at the planet's orbit if not for the disk's slightly sub-Keplerian motion, which shifts them inward.}, tend to circularize the planets. Massive planets, however, deplete these resonances by opening wide and deep gaps, enabling the farther-out, ``external'' eccentric Lindblad resonances to govern system evolution \citep[][]{Artymowicz93,Goldreich03,Sari04}. \cite{Duffell15} find that when non-accreting planets of ${\sim}1$ $\Mj$ are given an initial eccentricity, this mechanism can pump it to ${\sim}0.07$, of order their disk aspect ratio.

For multi-Jupiter-mass planets, the 1:3 eccentric Lindblad resonance allows disk eccentricity to grow as a form of instability \citep[e.g.,][]{Kley06}, and then be shared with the planet via secular back-reaction \citep[e.g.,][]{Papaloizou01,Bitsch13,Dunhill13,Ragusa18}. \citet{Papaloizou01} find that a 1 $\Mj$ planet does not become eccentric but 10 $\Mj$ does; \cite{Ragusa18} similarly find that the eccentricity of a non-accreting 13 $\Mj$ planet can grow to ${\gtrsim}$0.1; and \citet{Dunhill13} find that a 5 $\Mj$ planet still does not become eccentric, but note that this result should depend on disk parameters such as viscosity and temperature.

Planetary accretion can be expected to drain the circularizing, co-orbital resonances, facilitating eccentricity growth even for lower-mass planets. \cite{DAngelo06}, for instance, find that 2-3 $M_J$ planets can reach eccentricities of ${\gtrsim}0.1$; they achieve this using a low accretion rate and without adding the accreted mass to the planet's.\footnote{This is justified by their relatively low accretion rate and short simulation time.} Our work builds on theirs, but uses a higher accretion rate that grows the gas giant from its core mass. In this regime, we demonstrate that individual, accreting, initially circular super-Jupiters can acquire eccentricities of ${\gtrsim}0.25$. More importantly, we provide quantitative measurements of the gap depth and width for such a planet.

Taking these factors into account, we produce a self-consistent hydrodynamics simulation that tracks the accretion and orbital evolution of PDS~70b over 4.6 Myr (${\sim}\num{39000}$ orbits). Our fiducial model for the disk-planet system successfully reproduces the location and mass of the planet, along with the cavity size of the disk. \secref{sec:method} describes our hydrodynamical methods. \secref{sec:result} presents our model, \secref{sec:rad} shows how it compares with ALMA observations, and \secref{sec:conclude} concludes and discusses future directions.

\section{HYDRODYNAMICS SIMULATIONS}
\label{sec:method}
We use the graphics processing unit (GPU)-accelerated code \texttt{PEnGUIn} \citep{FungThesis15} to simulate disk-planet interaction in 2D. Apart from our accretion prescription, described below, the code is identical to that used for \citet{Fung18}, who similarly simulated planetary migration in disks.

In the present study, the planet accretes mass from each of the nearby cells at a rate of:\begin{subequations}\label{eq:acc_rate}
\begin{equation}
    \dot{\Sigma}_{\rm acc}(r, \phi) = -\frac{\Sigma(r, \phi)}{k~t_{\rm ff}} = -\frac{\Sigma(r, \phi)}{10 \sqrt{2 r_{\rm acc}^3/GM_p}}  \, ,
\end{equation}
limited spatially by a Gaussian function around the planet:
\begin{equation}
  \dot{\Sigma}_{\rm max}(r, \phi) = -\frac{\dot{M}_{\rm p,max}}{2\pi r_{\rm acc}^2} \exp{\left(-\frac{r^2 + r_p^2 - 2 r r_p \cos \phi'}{2r_{\rm acc}^2}\right)} \, ,
\end{equation}
where $\Sigma$ is the surface density of gas, $G$ the gravitational constant, $\Mp$ the mass of the planet,  $\dot{M}_{\rm p,max}$ the maximum accretion rate, $r_{\rm acc}$ the typical distance from the planet at which mass is accreted, $t_{\rm ff}$ the typical timescale for material at $r_{\rm acc}$ to free-fall onto the planet, and $k$ a constant that determines the accretion timescale.
We choose $r_{\rm acc} = r_{\rm H}/3$, where $r_{\rm H}=(M_{\rm p}/3\Mstar)^{1/3} r_{\rm p}$ is the planet's Hill radius, $\Mstar$ the star's mass, and $r_p$ the instantaneous distance between the planet and star.
For $k$, we find that $k=10$ gives a reasonable accretion rate that grows the planet to the size of PDS~70b over a duration comparable to the disk lifetime. Physically, this timescale is about 1.6 times longer than that given by Bondi accretion, which operates on a dynamical timescale $\sqrt{r_{\rm p}^3/G\Mstar}$.

We set $\dot{M}_{\rm p,max}$ to $100\Mj/$Myr, although in practice the rate is always lower. The actual planetary accretion rate may then be computed as
\begin{equation}
    \dot{M_p} = \int r\ dr\ d\phi \min(-\dot{\Sigma}_{\rm max}(r, \phi), -\dot{\Sigma}_{\rm acc}(r, \phi))
\end{equation}
integrated over the full domain.\footnote{Technically we accrete from the full domain, but the Gaussian function limits the maximum rate to about $1\Mj/$Myr outside 1 $\rH$, and $0.004\Mj/$Myr outside 1.5 $\rH$.}

While this prescription does not explicitly conserve angular momentum, the small $r_{\rm acc}$ ensures the material consumed co-orbits with the planet and has nearly the same specific angular momentum. We note that the planet's potential is softened with a smoothing length of $0.5$ times the local disk scale height, which is nearly always larger than $r_{\rm acc}$.
\end{subequations}

\begin{figure*}
    \centering
    \includegraphics[width=0.93\textwidth]{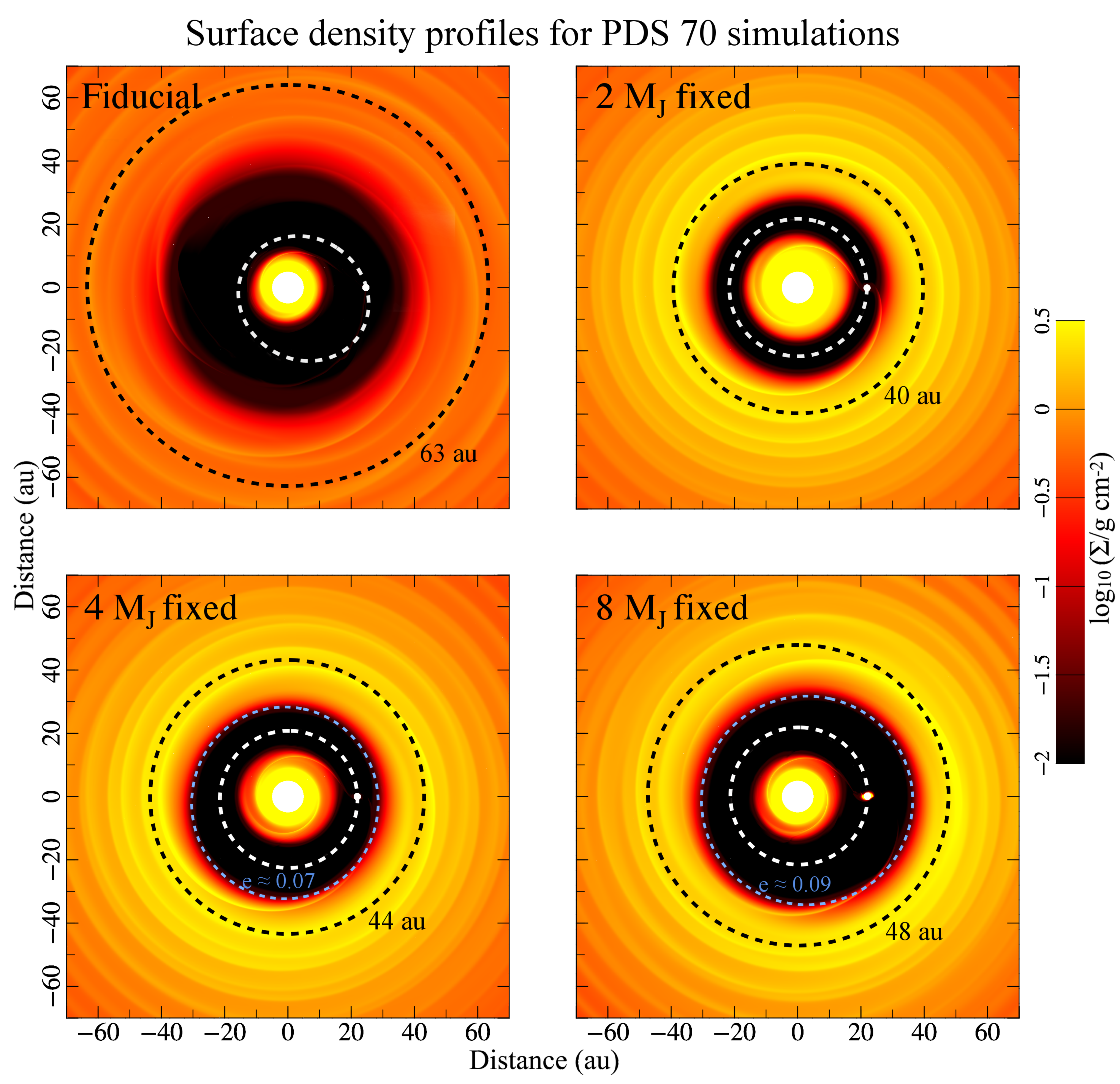}
    \caption{2D snapshots of gas surface density $\Sigma$ in our fiducial and control simulations. The fiducial snapshot is taken at 4.5 Myr, with the control snapshots taken at 250 kyr. Black dotted lines demarcate time-averaged, outer surface-density peaks; white lines the orbits of the planets; and light-blue lines the eccentric ``outer gap edge" where $\Sigma = \SI{0.01}{\gram\per\cm\squared}$.
    With eccentricity and accretion, the ${\sim}4\Mj$ planet in our fiducial model produces a wide gap that cannot be accounted for by a circularly orbiting, non-accreting companion.}
    \label{fig:2d_sigma}
\end{figure*}

\begin{figure*}
    \centering
    \includegraphics[width=0.9251\textwidth,,trim={0 0 0 0},clip]{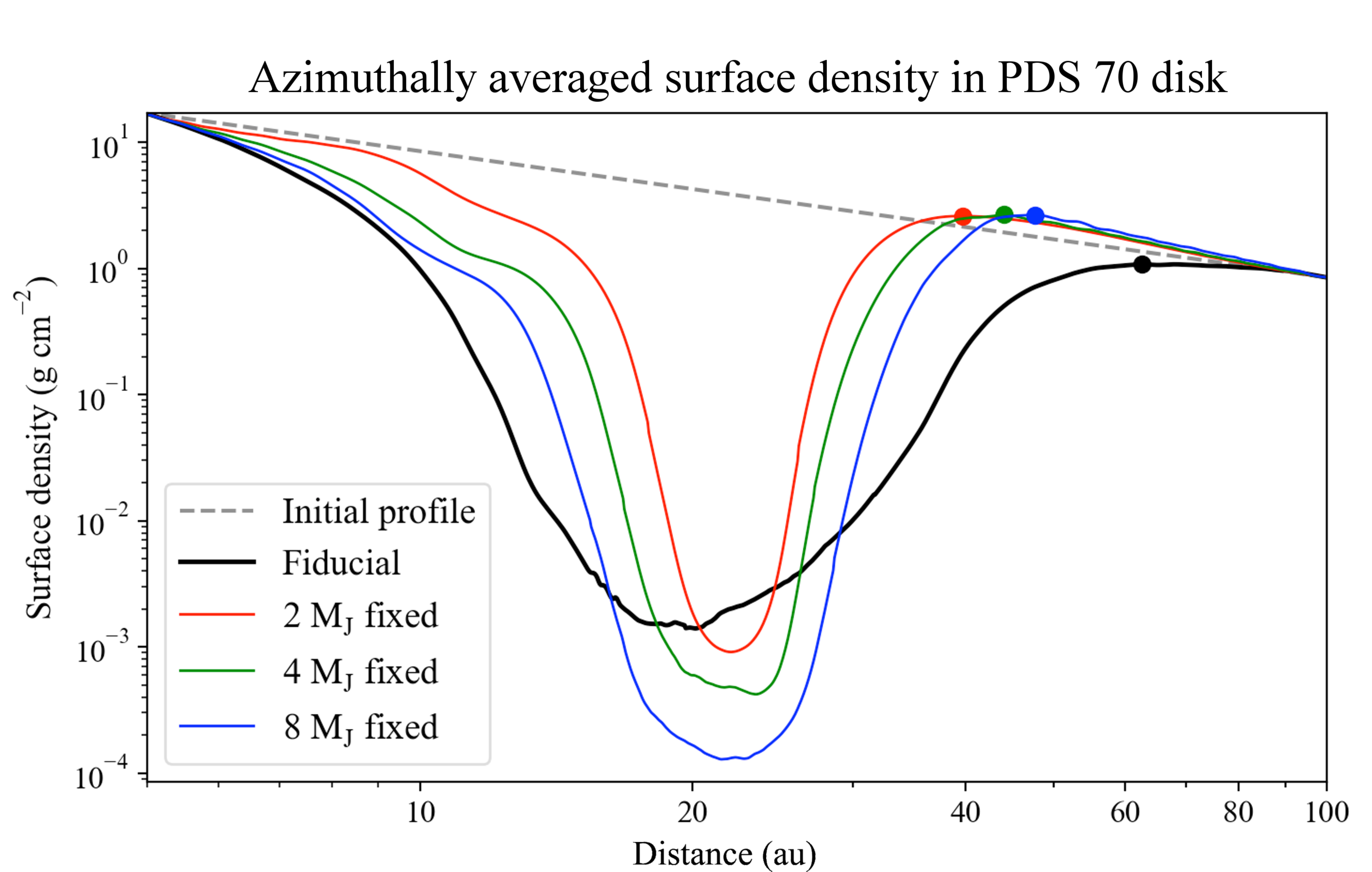}
    \caption{Azimuthally averaged counterparts for the 2D surface-density profiles of \figref{fig:2d_sigma}. We plot the the fiducial profile averaged over 4.5-4.6 Myr (black), by which time $\Mp \approx 4\Mj$, and the control profiles for 2, 4 and 8 $\Mj$ (red, green, and blue). We excise the regions near the planets $([r_p \pm 2 r_{\rm H}, \phi_p \pm 2 r_{\rm H}/r_p])$ to more clearly show the depth of the gaps. Dots indicate the local maximum --- where a dust ring may be present --- in the outer disk.}
    \label{fig:1d_averaged}
\end{figure*}
\subsection{Initial and Boundary Conditions}\label{sec:bc}
\label{sec:setup}
In line with the observations of \cite{Keppler18} and \cite{Long18}, we initialize our fiducial disk with a surface density profile of:
\begin{equation}
    \Sigma(r, \phi, t = 0) = \Sigma_0 \left(\frac{r}{22 \rm \ au}\right)^{-1}
\end{equation}
with $\Sigma_0 = \SI{3.864}{\gram\per\cm\squared}$. Based on calculations with the radiative-transfer code \texttt{DALI} \citep[][and our Section \ref{sec:rad}]{Bruderer13}, we impose a fixed sound-speed profile of
\begin{equation}\label{eq:sound_speed}
    c_s = c_0 \left(\frac{r}{22 \rm \ au}\right)^{-0.2}
\end{equation}
with $c_0 = \SI{0.35}{\km\per\second}$. This results in a flaring disk with aspect ratio $H \equiv h/r = 0.063(r/22 \rm \ au)^{0.3}$, and assuming a mean molecular weight of 2.34, a temperature profile given by $T = 35 \textup{ K}(r/22 \rm \ au)^{-0.4}$.

We set the notional disk viscosity parameter $\alpha_0 = 0.001$ \citep{ShakuraSunyaev1973}. This choice is motivated by the fact that PDS 70 has no clear large-scale asymmetry, implying that vortex formation at gap edges through the Rossby wave instability (RWI) \citep[e.g.,][]{Lovelace99, Li2000,LovelaceRomanova13} is suppressed. We enforce a steady state viscous disk for our initial condition, such that $\dot{M}_{\rm disk} \equiv 3\pi r v_{\rm r} \Sigma(t = 0)$ is constant and the radial velocity is
\begin{subequations}
\begin{equation}
    v_{\rm r}(r) = -\frac{3}{2}\frac{\nu(r)}{r} \, ,
\end{equation}
and the angular velocity is modified by pressure:
\begin{equation}
    \Omega = \sqrt{\Omega_{\rm K}^{2} + \frac{1}{r\Sigma}\frac{{\rm d}p}{{\rm d}r}} \, ,
\end{equation}
\end{subequations}
where $\nu = \alpha c_{\rm s} h$ is the kinematic viscosity, $\Omega_{\rm K} = \sqrt{GM_*/r^3}$ is the Keplerian orbital frequency. Per \cite{Keppler18}, we set $M_* = 0.76 \ M_\odot$. Since we require $\dot{M}_{\rm disk} \sim \alpha c_{\rm s} h \Sigma$ to be constant, we get that $\alpha = \alpha_0(r/\textup{au})^{-0.1}$, for our given disk profile.

Our simulation domain spans 5 to 100 au in radius and the full 2$\pi$ in azimuth. Radial boundaries are fixed to their initial values, presupposing that the simulated region of our disk lies within a viscous steady-state background. Similar fixed boundaries were used by \cite{DuffellDong15} and \cite{Keppler19} in their simulations of transition disks. To ensure numerical stability, the boundaries are softened using a standard wave-killing zone prescription \citep{valborro06}; in the present study, we damp all fields within 2 local scale heights of the boundaries to their initial values over 10 local orbital periods.

In our fiducial run, the planet is initialized as a 10 $\Me$ super-Earth, the typical core mass assumed for gas giants, on a 23 au circular orbit; over the first 100 Kyr of the simulation it migrates to 22 au, where it remains over the run's lifetime. We compare the results to three ``control'' runs, in which non-accreting planets of 2, 4, and 8 $\Mj$ are fixed on circular, 22 au orbits.
\subsection{Resolution}
\label{sec:res}

The grid dimensions used for the fiducial and control simulations are 960 ($r$) $\times$ 1944 ($\phi$) cells, spaced logarithmically in the radial and uniformly in azimuth; this corresponds to 20 cells per scale height at 22 au. In tests at both 50\% higher and lower than our fiducial resolution, as well as at different viscosities, we find that the position of the planet agrees to within 8\%.

\section{Results}
\label{sec:result}

We begin our investigation with a series of preliminary tests,varying parameters such as the planet's starting position ($20-60$ au), disk surface density ($\Sigma_0 =  \SIrange{4}{40}{\gram\per\cm\squared}$), and viscosity ($\alpha_0 = 0-2\times10^{-3}$). We also enable and disable mechanisms such as planetary migration and accretion. We use these models to narrow down the parameter space and identify relevant physical mechanisms. For example, our final choice of viscosity is motivated by them, as discussed in the previous section. In \secref{sec:acc}, we discuss the effects of changing disk surface density. As for the planet's starting location, we use it to investigate the effects of planetary migration on the gap structure.

Conceivably, a planet that starts closer to 60 au and over time migrates to 20 au may create a wider gap. Planets migrate due to angular momentum exchange with the disk (see \citealt{Kley12} for a review), and one way they can stop is by disk feedback \citep{Rafikov02}. Given a disk profile, we can control the ending position of a non-accreting planet by adjusting its mass \citep[e.g.,][]{Fung18}. We find that the effects of migration vary with viscosity: if $\alpha<10^{-3}$, the planet will create a wider gap, but the RWI will trigger at the gap edge and produce a large vortex; if $\alpha>10^{-3}$, viscosity will erase the planet's migration history over several thousand orbits and leave a significantly narrower gap than in PDS 70 \citep[e.g.,][]{Fung16}.

While our tests show that semimajor-axis evolution cannot reproduce the observed cavity, they do reveal that sufficiently massive planets tend to become eccentric. The resulting gaps are wider, but shallower, and without accretion, not clean enough to be optically thin in $^{12}$CO J=3--2 emission. The residual gap material torques the planet, and can restart planetary migration in ways that we do not fully understand. When we enable accretion, we find that in the limit of high accretion rate, the planet rapidly grows to the ``feedback'' mass (e.g., equation (1) of \citealt{Fung18}) and immediately begins to stall its initial migration. Moreover, the gap is substantially more depleted, which agrees with observations, and additionally helps stabilize the planet in its final position. In the end, we determine that a high accretion rate works best at reproducing PDS 70.

Our investigations lead to a best-fit fiducial model, in which the planet is free to both migrate and accrete, with $\alpha_0 = 0.001$ and $\Sigma_0 = \SI{3.86}{\gram\per\cm\squared}$. To accommodate the slow growth of the planet, we run it to 4.6 Myr (${\sim}\num{39000}$ orbits), approximately the current age of PDS~70 \citep{Muller18}. Accretion and eccentricity together enable the fiducial planet to assimilate a substantial fraction of the disk and produce a wide, deep cavity that resembles PDS~70, which would not be possible unless both mechanisms are accounted for.

\begin{figure*}
    \centering
    \includegraphics[width=0.93\textwidth]{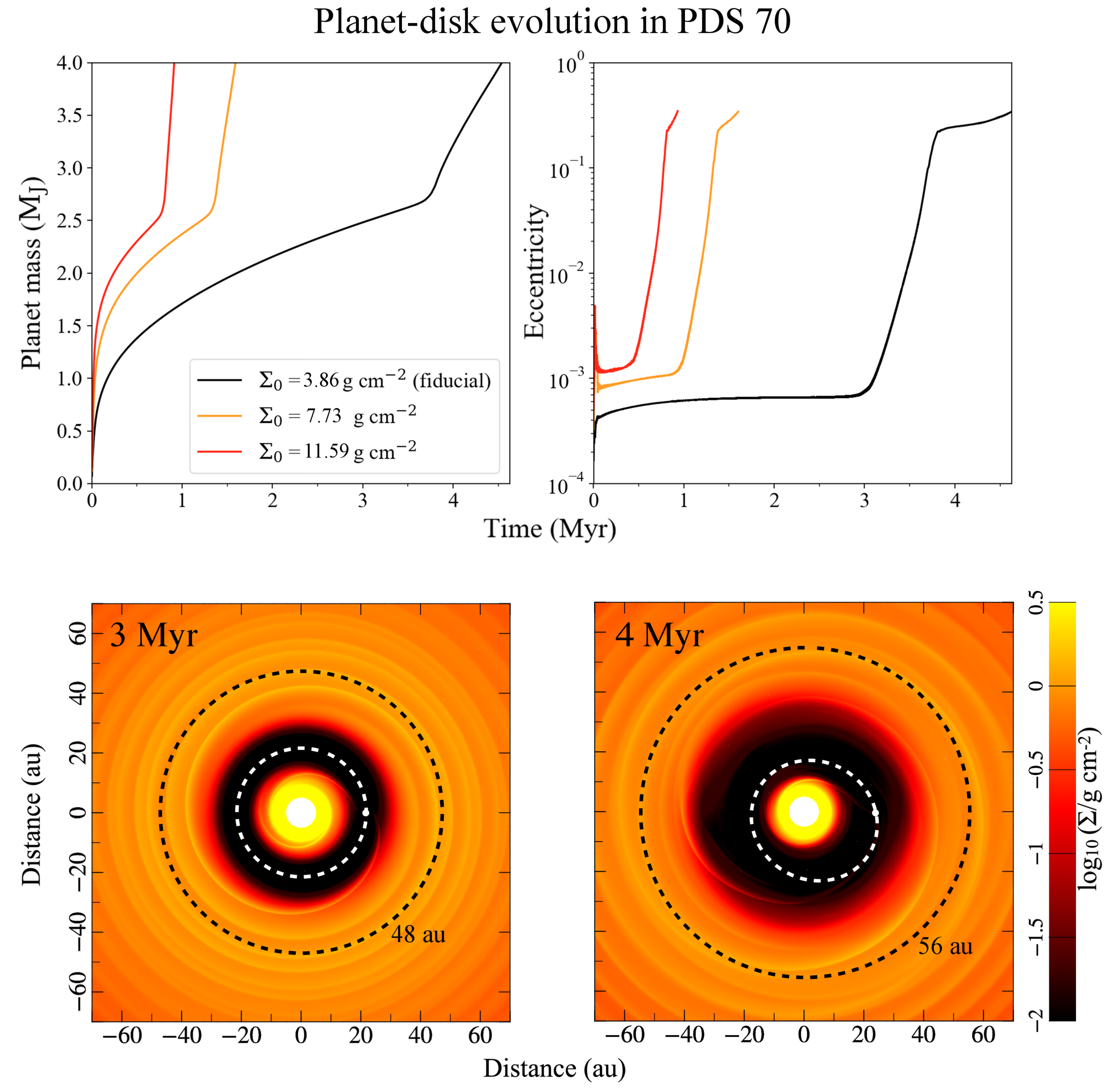}
    \caption{\textit{Top:} evolution of the planet as a function of disk density. In all cases, eccentricity grows exponentially when $M_p \gtrsim 2.5 \Mj$ ($q \gtrsim 0.003$), and saturates at $e_{\rm p} \approx 0.25$. Migration feedback stops all planets between 20-22 au.
    Eccentricity is plotted as a 20,000-timestep (${\sim}2000$ y) moving average. \textit{Bottom:} the fiducial disk at the start of exponential growth (left) and after saturation (right), demonstrating that eccentricity substantially widens the gap. }
    \label{fig:disk_mass}
\end{figure*}

In figures \ref{fig:2d_sigma} and \ref{fig:1d_averaged}, we compare the fiducial run to three representative control simulations in which planet masses and orbits are fixed; these are run to 250 Kyr (${\sim}\num{2000}$ orbits), by which time they should have reached steady-state \citep[e.g.,][]{Fung16}. Our control setups are similar to those used in \cite{Keppler19}, and likewise, do not reproduce the wide, observed gap. In \figref{fig:disk_mass} and \secref{sec:acc}, we discuss our ``comparison runs'', identical to our fiducial simulation except with greater surface density. The \hyperref[sec:appendix]{Appendix} presents our larger set of accreting, migrating, long-running simulations with $\alpha_0 = 0.001$, which admit quantitative comparison to our fiducial run.

By the end of the fiducial simulation, the planet has depleted the gas below $\Sigma = \SI{0.01}{\gram\per\cm\squared}$ out to 30 au, and produced a pressure peak at 62 au. Observationally, these would manifest as a ${\sim}$30 au optically thin cavity in $^{12}$CO and a ${\sim}60$ au ring in dust continuum, which we verify with radiative transfer modeling in Section \ref{sec:rad}. 


\subsection{Disk-Planet Evolution}
\label{sec:ecc}
In agreement with the results of \cite{Kley06}, we find in all of our simulations that the disk becomes eccentric when the planet-to-star mass ratio $q \gtrsim0.003$ (${\gtrsim}2.5 \Mj$ for PDS~70). This is seen in the 4 and 8 $\Mj$ control runs, where the disk remains visibly eccentric to the end (\figref{fig:2d_sigma}). It also occurs in the fiducial run when the planet grows to ${\sim}2.5~\Mj$, but the subsequent evolution differs because the planet experiences a back-reaction from the disk.

At the start of our fiducial simulation, the planet rapidly accretes at a rate of ${\sim} 10 \Mj /\rm Myr$, which decreases as the gap widens and deepens. By 3 Myr, when the planet reaches a mass of ${\sim} 2 \ \Mj$, the disk's outer surface density peak is at 48 au---${\sim}20$\% farther than in the comparable $2 \ \Mj$ control simulation---and the gap is depleted by a factor of ${\sim} 10^4$. Throughout this period, the planet's eccentricity $e_{\rm p} \lesssim 10^{-3}$, and its semimajor axis $a_{\rm p}$ is an essentially constant 21.7 au.

Past 3 Myr, the planet grows to $q \gtrsim 0.003$. The gap is now wide enough for external eccentric Lindblad resonances to strongly influence disk-planet interaction, making the disk eccentric \citep[e.g.,][]{Papaloizou01,Kley06}. Secular interactions subsequently drive exponential growth in the planet's eccentricity, which eventually saturates at $e_{\rm p} \approx 0.25$. This has two major consequences---first, it distributes the planet's torque over a larger radial range, making the gap substantially wider, albeit modestly shallower. Second, it enables the planet to access and consume more disk material, further enlarging the gap; see Figure \ref{fig:1d_averaged} for a comparison. Outside the now ${\gtrsim}60$ au-wide cavity, the disk is now far enough from the planet's gravity to remain essentially circular.

We obtain a higher planetary eccentricity than in previous studies of super-Jupiters in disks. This is likely caused by our longer simulations ($3.9\times10^4$ orbits) and a higher accretion rate. \citet{DAngelo06}, for instance, simulated up to 7000 orbits and had an accretion rate about 10 times lower than ours; while \citet{Ragusa18} exceeded our simulation time in number of orbits, but had no accretion. Accretion, in particular, may be key to determining the planet's eccentric evolution. \citet{Duffell15} proposed that when the planet collides with the edge of its gap, it replenishes the co-orbital Lindblad resonances and causes the eccentricity to damp. If, as in our fiducial model, the planet accretes the gap-edge material, this damping effect is nullified and eccentricity could potentially grow much higher.

Our model includes many simplifying assumptions. First, our choice of fixed inner and outer boundaries imply that our simulation domain lies within an unchanging steady-state viscous disk. In reality, over Myr timescales, the underlying disk mass should disperse due to mechanisms such as viscous evolution and photoevaporation. Disk dispersal could be modeled with a time-dependent density at our boundaries, but \textit{a priori}, it is unclear what the most realistic prescription would be, so we choose not to do implement one. As a result, the way the planet grew in the first 3 Myr of our fiducial simulation is somewhat lacking in realism. This would have an impact on our results if the planet's eccentricity evolution 1) depends on its accretion history, or 2) occurs over a timescale as long as the disk dispersal time of roughly 3 Myr. We examine this more closely in \secref{sec:acc} by repeating our fiducial model using different disk surface densities, corresponding to different stages of the disk's evolution, and observing the effects on the planet's eccentricity. 

Second, our simplistic accretion prescription has much room for improvement. The present model has no end-point---the planet will always accrete as long as there is material nearby. In reality, planetary growth can stop if accretional heating is able to counterbalance radiative cooling of the planet's envelope \citep[e.g.,][]{Lee15,Ginzburg16}, or if a circumplanetary disk forms and accretion becomes restricted by angular-momentum transfer. At the same time, our assumption of a steady-state background viscous disk provides the planet with a never-ending supply of gas, even though real disks are finite. Given these limitations, we run our simulations only until planetary eccentricity saturates, and do not aim to model the ultimate fate of the planet.

\begin{figure*}[!ht]
\includegraphics[width=\textwidth]{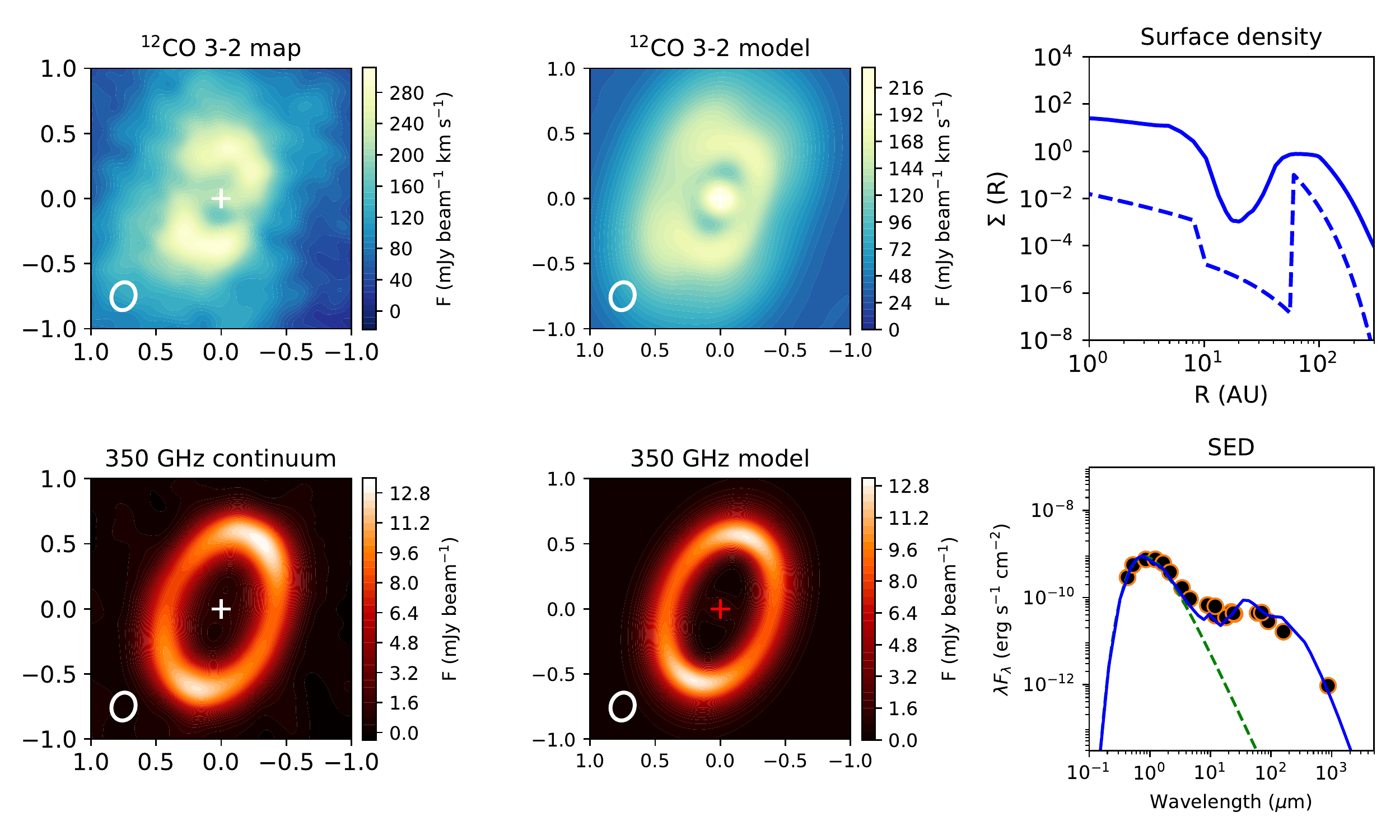}
\caption{\texttt{DALI} model of $^{12}$CO J=3--2 and 350 GHz continuum emission of PDS 70, using the fiducial simulation for gas surface density. Left panels show observations, middle panels model images, the top right the gas (solid) and dust (dashed) profiles, and the bottom right the observed SED with a model overlaid. Beam profiles are indicated at the bottom-left of each image; models are ray-traced at 113 pc \citep{GaiaDR2}. The continuum and SED models reproduce the observations; the $^{12}$CO model map has the correct morphology, but shows a strong excess of emission in the inner disk.}
\label{fig:dalimodel}
\end{figure*}

\subsection{Accretion and Disk Mass}
\label{sec:acc}
Disk evolution in our fiducial model is qualitatively consistent over a wide range of disk masses. We demonstrate this using two runs with identical initial conditions, except with 2$\times$ and 3$\times$ the surface density and a lower resolution, 634 ($r$) $\times$ 1296 ($\phi$). These low-resolution runs reproduce the timescale for eccentricity excitation to within $\lesssim$10\% of that at the fiducial resolution. As shown in \figref{fig:disk_mass}, planetary mass and eccentricity evolve similarly to our fiducial run, but the increased disk mass means that the planet can more rapidly reach the critical $q \gtrsim 0.003$ for eccentricity excitation. In particular, this threshold does not show a clear dependence on the disk mass. The characteristic timescale for eccentricity growth is about a few hundred thousand years, roughly 10 times shorter than typical disk-dispersal times (${\sim}3$ Myr). Therefore, disk depletion physics, or equivalently, the age and accretion history of the planetary core, should not play a major role
in the eccentricity evolution of the planet, but future investigations coupling disk physics with disk-planet interaction are needed to fully address this.

Over Myr timescales, a lower disk mass may help stabilize the system. When the initial disk mass is of order a few $\Mj$, as in our fiducial run, the formation of a giant planet would remove enough mass to severely inhibit migration via disk-planet interaction. If the initial disk mass is too high, or the planetary accretion rate is too low, migration could move the planet across the entire disk on very short timescales. Indeed, we have observed such fast migration in some of our tests with non-accreting, super-Jupiter planets.

Finally, we note that in our fiducial model, the total gas mass within 100 au is $\sim 6 \Mj$ (13 and 25 $\Mj$ for the higher density models). Even though there is a continuous inflow of gas from the outer boundary, it may still be insufficient to plausibly form a super-Jupiter planet. The fact that it does form in our simulations is again due to our highly efficient, simplistic accretion prescription, discussed in the previous section. For this reason, we caution the reader that our disk models are not fully indicative of the formation conditions of PDS 70b.

\section{Radiative transfer modeling}
\label{sec:rad}
Using \texttt{DALI}, we model the $^{12}$CO J=3--2 and 350 GHz dust continuum emission of our fiducial simulation, and compare with ALMA observations. \texttt{DALI} is a physical-chemical modeling code \citep{Bruderer13} which self-consistently solves for gas temperature, molecular abundances, and excitation given gas and dust surface density profiles. For the gas surface density, we use the output of the fiducial run. For dust, we assume a power law with an exponential tail:
\begin{equation}\label{eq:dust_sigma}
\Sigma_{\rm d}(r) = \Sigma_{\textup{d}, c} \left(\frac{r}{r_c}\right)^{-\gamma} {\rm exp}\left(-\left(\frac{r}{r_c}\right)^{2-\gamma}\right)
\end{equation}
with $\Sigma_{\textup{d}, c}=\SI{0.8}{\gram\per\cm\squared}$, $r_c = 15 \rm \ au$ and $\gamma = 1$, with the surface density reduced by a factor of 10$^6$ for $r < r_{\rm cav}$, and 10$^4$ for $r<$ 10 au to reproduce the inner dust disk. From the pressure peak in our fiducial run, we set $r_{\rm cav} =60 \rm \ au$ (see Figure \ref{fig:dalimodel}). We extrapolate the dust profile for $r>100 \rm \ au$ with equation \eqref{eq:dust_sigma}, and the gas profile by a piecewise version, choosing coefficients in the inner and outer disk that ensure continuity; this yields an overall gas-to-dust ratio of 73. More details on this procedure are provided in \citet{vanderMarel16}.

We have reduced the ALMA Band 7 observations of \citet{Long18} of the 350 GHz continuum and $^{12}$CO 3--2 emission using the default pipeline scripts, and cleaned the data using natural weighting, resulting in a beam size of 0.22''$\times$0.18''. Figure \ref{fig:dalimodel} shows that the SED and continuum ring closely match the data. The $^{12}$CO emission model shows a similar morphology to that observed---indicating that it is optically thin at the planet's location---and has a comparable integrated flux. However, it also exhibits a bright central peak, significantly stronger than that found by ALMA---likely the result of our inner-disk boundary conditions. We address this more thoroughly in the following section (\secref{sec:conclude}).

\section{Conclusions and Discussions}
\label{sec:conclude}
Using 2D hydrodynamics simulations and radiative transfer post-processing, we demonstrate that the super-Jupiter PDS~70b, located 22 au from its host star, is by itself capable of carving out the ${\sim}30$ au gas and ${\sim}60$ au dust cavity in which it resides (Figures \ref{fig:2d_sigma} and \ref{fig:1d_averaged}). Radiative transfer modeling yields results consistent with the observed continuum emission and SED of PDS~70, as well as the overall disk morphology in $^{12}$CO (Figure \ref{fig:dalimodel}).

Our model makes the following predictions about PDS 70:
\begin{enumerate}
\item PDS~70b is on an eccentric orbit with $e_{\rm p}\gtrsim0.2$. This naturally results from 
disk-planet interaction
and is required to explain the width of the observed cavity.

\item The planet grew {\it in situ}; gas accretion in combination with disk feedback is effective at inhibiting planet migration.

\item To produce a wide, clean gap, the planet must have consumed a large region of the disk as it evolved.
\end{enumerate}

While we do not model the scattered-light emission of the PDS 70 disk, it is possible to make rough estimates. If we assume the gas-to-dust ratio is 73 (see \secref{sec:rad}), and that 3.2\% of the dust comprises of micron-size grains \citep{Keppler18} with a scattering opacity of about $\SI{3e3}{\cm\squared\per\gram}$ in the $\mu$m range \citep[e.g.,][]{Tazaki18}, then the optically thin region extends to roughly 49 au in our fiducial model. This is encouraging, considering that \citet{Keppler18} measured the outer scattered-light gap edge to be at ${\sim}54 \rm \ au$, but a quantitative comparison would require consideration of factors such as grain drift and growth \citep[e.g.,][]{Birnstiel12,Dong12,Zhu12}, as well as instrumental effects including noise and resolution.

One alternative to the single-planet scenario would be for PDS~70b to occupy a circular orbit, with additional, unseen planets at larger radii carving out the observed cavity \citep{DodsonSalyk11,Zhu2011,DuffellDong15,Keppler19}. Given PDS~70b's mass, however, our simulations predict that it has likely become eccentric---suppressing this would require a disk surface density an order of magnitude or more lower than that observed. Thus, while additional planets may exist alongside PDS~70b, they are not required to produce the transition cavity in the system.

Our fiducial model has several areas for improvement. For instance, the optically thick $^{12}$CO emission it predicts from the inner disk far exceeds observations, implying that the simulated inner disk---approximately 10 au in radius---is likely too large. This size is influenced by the location of the inner hydrodynamical boundary, which we have set to 5 au because this work focuses on large-scale ring and gap, rather than closer-in regions. Additionally, the details of gas-grain interaction and vertical stratification in the inner disk could significantly elevate the temperature and emission of $^{12}$CO in our fiducial model as compared to that in the real PDS~70. Testing a closer inner boundary and alternative vertical structures would improve our fit of the inner disk, but at a substantially elevated computational cost.

With our current planetary-accretion prescription, the planet simply consumes the material in its vicinity, subject to an artificial maximum rate. Including thermal physics and circumplanetary-disk dynamics would better constrain the growth rate of the planet, and thus the timescale on which the disk-planet system evolves. Such changes, however, would not alter our basic conclusion---that a sufficiently massive super-Jupiter would spontaneously go eccentric, and consume a large region of the disk to create a wide, deep cavity.

Another concern is that our simulations are in 2D. Running our 4.6 Myr long fiducial model in 3D would currently be impractical, requiring years in wall-clock time. 2D may be sufficient --- previous work has shown that 3D gap-opening \citep{Fung17a} and planetary torques \citep{Fung17b} are reasonably well-captured in 2D; however, it remains to be seen how closely 2D eccentricity excitation and planetary accretion would match their 3D counterparts. 2D also cannot address whether the orbit of PDS~70b is inclined, and the effects this would have on the cavity structure. In the future, detailed studies of the 3D dynamics would improve understanding of how gas giants and transition disks form.

Single, super-Jupiter giants naturally carve out wide transition cavities, with no need to fine-tune free parameters. Thus, future surveys may find it worthwhile to investigate the opposite question---what fraction of transition disks are sculpted by giant planets \citep{Dong16}? Besides disk morphology, our findings also raise interest in the fact that super-Jupiters themselves are observed to be highly eccentric \citep{Butler06}. Our simulations show that disk-planet interaction may be a cause, and we intend to study this more thoroughly in a forthcoming paper.

\acknowledgements
{\small \noindent \\We thank Jaehan Bae, Eugene Chiang, Ruobing Dong, Kaitlin Kratter, Cristobal Petrovich, Jonathan Williams, and Zhaohuan Zhu for useful suggestions and discussions, as well as the anonymous referees for providing helpful comments that improved our manuscript. This collaboration was initiated at the Aspen Center for Physics, which is supported by National Science Foundation grant PHY-1607611. This work was performed under contract with the Jet Propulsion Laboratory (JPL) funded by NASA through the Sagan Fellowship Program executed by the NASA Exoplanet Science Institute. This paper makes use of the following ALMA data: ADS/JAO.ALMA/2015.1.00888.S. ALMA is a partnership of ESO (representing its member states), NSF (USA) and NINS (Japan), together with NRC (Canada) and NSC and ASIAA (Taiwan), in cooperation with the Republic of Chile. The Joint ALMA Observatory is operated by ESO, AUI/NRAO and NAOJ.}

\bibliographystyle{yahapj}
\bibliography{Lit}
\appendix 
\label{sec:appendix}
Below we present results from simulations qualitatively similar to our fiducial run. Specifically, we select those runs that include planetary migration and accretion, use an $\alpha_0 = 0.001$ to suppress the RWI, and have a duration of at least 100 kyr. This accounts for 29 simulations. For reference, we also include the results of our fiducial (green), comparison (blue), and control runs (red).

In the table below, the $R_{\Sigma}$ of a given simulation is the ratio between its initial disk mass (or surface density) versus that in the fiducial simulation. $\dot{M}_{\rm p, max}$ is defined in \secref{sec:method}, while $r_{\rm p, 0}$ represents the starting position of the planet. The ``stopping point'' of migration is given by $r_{\rm p, end}$.

Eccentricity evolves throughout the duration of our simulations, making a direct measurement challenging. We estimate the cutoff mass for eccentricity onset, $M_{\rm p, ecc}$, as that when $e_{\rm p}$ equals the local disk aspect ratio $h/r$. We find eccentricity excitation in almost all runs of sufficient duration, with the expected critical planet-to-star mass ratio for excitation $q_{\rm ecc} \sim 0.003$. We note that our estimator is a conservative one: it does not indicate eccentricity at all if a simulation is terminated before $e_{\rm p}$ has had a chance to grow to large values.

\fontsize{8}{10}\selectfont\centering
\begin{longtable*}{|l|l|l|l|l|l|l|l|l|}
\hline
Resolution & Boundaries (au) & $R_{\Sigma}$ & $\dot{M}_{\rm p, max}$ ($\Mj$/Myr) & Duration (y) & $r_{\rm p, 0}$ (au) & $r_{\rm p, end}$ (au) & $M_{\rm p, ecc}$ ($\Mj$) & Notes         \\ \hline
\rowcolor[HTML]{C2F2C1} 960 $\times$ 1944   & 5--100      & 1              & 100       & \num[round-mode=places,round-precision=1,scientific-notation=true]{4600000}     & 23                    & 21                     & 2.7            & Fiducial      \\
\rowcolor[HTML]{B9E3E5}634 $\times$ 1296   & 5--100      & 2              & 100       & \num[round-mode=places,round-precision=1,scientific-notation=true]{1603486}     & 23                    & 22                     & 2.6            & 2$\times$ comparison \\
\rowcolor[HTML]{B9E3E5}634 $\times$ 1296   & 5--100      & 3              & 100       & \num[round-mode=places,round-precision=1,scientific-notation=true]{932793}      & 23                    & 20                     & 2.5            & 3$\times$ comparison \\
\rowcolor[HTML]{F4C2BF}960 $\times$ 1944   & 5--100      & 1              & ---         & \num[round-mode=places,round-precision=1,scientific-notation=true]{250000}      & 22                    & 22                        & ---                & 2 $\Mj$ fixed  \\
\rowcolor[HTML]{F4C2BF}960 $\times$ 1944   & 5--100      & 1              & ---         & \num[round-mode=places,round-precision=1,scientific-notation=true]{250000}      & 22                    & 22                        & ---                & 4 $\Mj$ fixed  \\
\rowcolor[HTML]{F4C2BF}960 $\times$ 1944   & 5--100      & 1              & ---         & \num[round-mode=places,round-precision=1,scientific-notation=true]{250000}      & 22                    & 22                        & ---                & 8 $\Mj$ fixed  \\
720 $\times$ 1296   & 5--150      & 1.5            & 10        & \num[round-mode=places,round-precision=1,scientific-notation=true]{2785941}     & 30                    & 25                     & 2.6            &               \\
720 $\times$ 1296   & 5--150      & 2.6            & 50        & \num[round-mode=places,round-precision=1,scientific-notation=true]{1000000}     & 30                    & 26                      & 2.7            &               \\
720 $\times$ 1296   & 5--150      & 2.5            & 75        & \num[round-mode=places,round-precision=1,scientific-notation=true]{829611}      & 25                    & 20                     & 2.3            &               \\
720 $\times$ 1296   & 5--150      & 2.9            & 50        & \num[round-mode=places,round-precision=1,scientific-notation=true]{791186}      & 25                    & 17                     & 2.1            &               \\
720 $\times$ 1296   & 5--150      & 2.6            & 400       & \num[round-mode=places,round-precision=1,scientific-notation=true]{660935}      & 25                    & 23                     & 2.6            &               \\
634 $\times$ 1296   & 5--100      & 4            & 100       & \num[round-mode=places,round-precision=1,scientific-notation=true]{648257}      & 23                    & 20                     & 2.6            &               \\
720 $\times$ 1296   & 5--150      & 2.5            & 100       & \num[round-mode=places,round-precision=1,scientific-notation=true]{639526}      & 25                    & 21                     & 2.3             &               \\
975 $\times$ 1296   & 1--100      & 4              & 100       & \num[round-mode=places,round-precision=1,scientific-notation=true]{606078}      & 23                    & 19                     & ---                &               \\
960 $\times$ 1944   & 5--100      & 4              & 100       & \num[round-mode=places,round-precision=1,scientific-notation=true]{576143}      & 23                    & 19                     & 2.5            &               \\
720 $\times$ 1296   & 5--150      & 2              & 400       & \num[round-mode=places,round-precision=1,scientific-notation=true]{466995}      & 25                    & 24                      & ---                &               \\
720 $\times$ 1296   & 5--100      & 2              & 100       & \num[round-mode=places,round-precision=1,scientific-notation=true]{465950}      & 23                    & 22                      & ---                &               \\
720 $\times$ 1296   & 5--150      & 2.5            & 50        & \num[round-mode=places,round-precision=1,scientific-notation=true]{458646}      & 25                    & 20                     & ---                &               \\
720 $\times$ 1296   & 5--150      & 3.5            & 50        & \num[round-mode=places,round-precision=1,scientific-notation=true]{425602}      & 30                    & 24                     & 2.4             &               \\
720 $\times$ 1296   & 5--150      & 3              & 50        & \num[round-mode=places,round-precision=1,scientific-notation=true]{411381}      & 25                    & 19                     & 1.9            &               \\
720 $\times$ 1296   & 5--150      & 2              & 50        & \num[round-mode=places,round-precision=1,scientific-notation=true]{384290}      & 25                    & 22                      & ---                &               \\
720 $\times$ 1296   & 5--150      & 3.6            & 400       & \num[round-mode=places,round-precision=1,scientific-notation=true]{339262}      & 30                    & 27                     & 2.8            &               \\
720 $\times$ 1296   & 5--150      & 3.2            & 400       & \num[round-mode=places,round-precision=1,scientific-notation=true]{273416}      & 25                    & 22                        & ---                &               \\
720 $\times$ 1296   & 5--150      & 2              & 5         & \num[round-mode=places,round-precision=1,scientific-notation=true]{263231}      & 55                    & 24                     & ---                &               \\
720 $\times$ 1296   & 5--150      & 2.8            & 20        & \num[round-mode=places,round-precision=1,scientific-notation=true]{263044}      & 55                    & 27                     & ---                &               \\
720 $\times$ 1296   & 5--150     & 2.4            & 10        & \num[round-mode=places,round-precision=1,scientific-notation=true]{220608}      & 55                    & 24                     & ---                &               \\
720 $\times$ 1296   & 5--150      & 5.2            & 200       & \num[round-mode=places,round-precision=1,scientific-notation=true]{215877}      & 50                    & 22                     & 3.2            &               \\
720 $\times$ 1296   & 5--150      & 1.7            & 5         & \num[round-mode=places,round-precision=1,scientific-notation=true]{213419}      & 55                    & 31                      & ---                &               \\
720 $\times$ 1296   & 5--150      & 3.6            & 200       & \num[round-mode=places,round-precision=1,scientific-notation=true]{184703}      & 25                    & 22                     & ---                &               \\
720 $\times$ 1296   & 5--150      & 4.8            & 150       & \num[round-mode=places,round-precision=1,scientific-notation=true]{183626}      & 50                    & 23                     & 3.0             &               \\
720 $\times$ 1296   & 5--150      & 4.6            & 100       & \num[round-mode=places,round-precision=1,scientific-notation=true]{161558}      & 50                    & 20                     & 2.5             &               \\
720 $\times$ 1296   & 5--150      & 3.2            & 25        & \num[round-mode=places,round-precision=1,scientific-notation=true]{148587}      & 40                    & 22                     & ---                &               \\
720 $\times$ 1296   & 5--150      & 5              & 400       & \num[round-mode=places,round-precision=1,scientific-notation=true]{122338}      & 25                    & 22                     & 2.5            &               \\
720 $\times$ 1296   & 5--150      & 5              & 200       & \num[round-mode=places,round-precision=1,scientific-notation=true]{119844}      & 50                    & 24                      & 3.3            &               \\
720 $\times$ 1296   & 5--150      & 5              & 100       & \num[round-mode=places,round-precision=1,scientific-notation=true]{109457}      & 30                    & 20                     & 2.2            &               \\

 \hline
\end{longtable*}

\end{document}